\documentclass[prb,
twocolumn,
superscriptaddress,showpacs,amsmath,amssymb]{revtex4}

\begin{document}

\title{Double Hawking temperature: from black hole to de Sitter}

\author{G.E.~Volovik}
\affiliation{Low Temperature Laboratory, Aalto University,  P.O. Box 15100, FI-00076 Aalto, Finland}
\affiliation{Landau Institute for Theoretical Physics, acad. Semyonov av., 1a, 142432,
Chernogolovka, Russia}

\date{\today}

\begin{abstract}
The double Hawking temperature $T=2T_H$ appears in some approaches to the Hawking radiation, when the radiation is  considered in terms of the quantum tunneling. We consider the origin of such temperature for the black hole horizon and also for the cosmological horizon in de Sitter spacetime. In case the black hole horizon, there are two contributions to the tunneling process of radiation, each being governed by the   temperature $T=2T_H$. These processes are coherently combined to produce the radiation with the Hawking temperature $T_H$.  This can be traditionally interpreted as the pair creation of two entangled particles, of which one goes towards the centre of the black hole, while the other one escapes from the black hole. In case of the cosmological horizon, the  temperature $T=2T_H$ is physical. While the creation of the entangled pair is described by the Hawking temperature, the de Sitter spacetime allows for the another process, in which only  single (non-entangled) particle inside the cosmological horizon is created. This process is characterized by the local temperature  $T=2T_H$. The local single particle process takes place also outside the black hole horizon, but it is exponentially suppressed.
\end{abstract}
\pacs{
}

\maketitle

 \tableofcontents
% \newpage

%\twocolumn

\section{Introduction}

The double Hawking temperature $2T_H$ appears in some approaches to the Hawking radiation from the black hole and cosmological horizons considered as quantum tunneling.
In the case of the black hole horizon, the $2T_H$ is the result of the incorrect choice of reference frame, while the correct choice (the Painleve-Gullstrand coordinate system) gives the Hawking temperature $T_H$.
In the de Sitter Universe the situation is different, the temperature $T=2T_H$ is physical: it is the local temperature experienced by matter well inside the cosmological horizon. We discuss the possible connections between these two manifestations of the double Hawking temperature.

\section{$2T_H$ problem for black holes}
\label{2THblackhole}

The quantum tunneling approach allows us to study different processes without consideration of the details of the microscopic quantum field theory. The Hawking radiation from the black hole\cite{Hawking1975} is an example when the complicated process is highly simplified by the method of quantum tunneling,\cite{Volovik1999,Wilczek2000,Volovik2009} 
The Hawking temperature is obtained when the tunneling rate is compared with the Boltzmann factor.
 In this approach
the Painleve-Gullstrand (PG) coordinate system \cite{Painleve,Gullstrand} is used with the metric:
 \begin{equation}
ds^2= - dt^2(1-{\bf v}^2) - 2dt\, d{\bf r}\cdot {\bf v} + d{\bf r}^2 \,,
\label{PGmetric}
\end{equation}
where $v^2(r) =R/r$; $R=2M$ is the position of the black hole horizon ($G=c=\hbar=1$); the contracting snd expanding velocities ${\bf v}$ correspond to the black hole and white hole correspondingly (for extension of the PG metric to the black and white holes with several horizons see Ref. \cite{Volovik2022b}).
This metric does not have singularity at the horizon, and for the black hole realization it gives the tunneling exponent corresponding to Hawking temperature $T_H= 1/8\pi M$. This process can be interpreted as the pair production of two entangled particles, of which one goes towards the center of the black hole, while other one escapes.

In some approaches to the Hawking radiation as quantum tunneling, there exists the $2T_H$ problem.\cite{Hooft1984,Akhmedov2006,Mitra2007,Laxmi2021} We consider this problem using Klein–Gordon equation  for massive field
in a curved background,\cite{Akhmedov2006} 
which leads to the relativistic Hamilton--Jacobi equation for the classical action:
 \begin{equation}
g^{\mu\nu}\partial_\mu S \partial_\nu S + m^2 =0\,.
\label{HJ}
\end{equation}
Let us start with the PG metric (\ref{PGmetric}), since it does not have coordinate singularity at the horizon.
One has for the fixed energy $E$:\cite{Akhmedov2006} 
 \begin{equation}
-E^2 +(1-v^2) \left(\frac{dS}{dr}\right)^2 + 2vE\frac{dS}{dr}+m^2=0\,.
\label{KG}
\end{equation}
This gives for the classical action
 \begin{equation}
S=-\int dr \frac{Ev}{1-v^2} \pm \int  \frac{dr}{1-v^2}\sqrt{E^2-m^2(1-v^2)} \,.
\label{SPG}
\end{equation}
For the plus sign in the second term the imaginary parts of two terms cancel each other. This corresponds to the incoming trajectory of the particle.
For the minus sign in the second term the total tunneling exponent describes the radiation with the Hawking temperature $T_H$:
\begin{eqnarray}
 \exp{ \left(-2\,{\rm Im}\, S\right)}= \exp{ \left(-\frac{E}{2T_H} \right)} \exp{ \left(-\frac{E}{2T_H} \right)} =
 \nonumber
% \label{TH1}
 \\
 =\exp{ \left(-\frac{E}{T_H} \right)} \,.
\label{TH2}
\end{eqnarray}
Note that each of the two terms in Eq.(\ref{SPG}) corresponds to the effective temperature $2T_H$. 
That is why if one of the two terms is lost in calculations, the temperature $2T_H$ erroneously emerges.
The similar product of the terms with $2T_H$ each has been obtained also for the black holes with several horizons.\cite{Singha2022}

In Ref. \cite{Akhmedov2006}  the Schwarschild coordinates were also used (see also Ref.\cite{Srinivasan1999}):
\begin{equation}
ds^2=- \left(  1- \frac{R}{r} \right)dt^2 + \frac{dr^2}{1- \frac{R}{r} } + r^2 d\Omega^2
\,.
\label{StaticMetric}
\end{equation}
This gives only the second term in Eq.(\ref{SPG}), and as a result the $2T_H$ temperature is obtained.
In Refs.\cite{Srinivasan1999,Mitra2007} it was argued that the Hawking temperature can be restored by consideration of the balance between the emission $p_{\rm em} \propto \exp(-E/2T_H)$, which comes from the minus sign,  and the absorption $p_{\rm abs} \propto \exp(E/2T_H)$, which comes from the plus sign.  Then the ratio $p_{\rm em}/ p_{\rm abs}$ gives the Hawking temperature.

However, the consideration of the exponential absorption is somewhat unnatural. The more natural view is that the first term in Eq.(\ref{SPG}) can be restored  if one takes into account, that the Schwarschild metric  has singularity at the horizon. As a result, the transition from PG coordinates to  Schwarschild coordinates requires the singular coordinate transformation:  
\begin{equation}
dt \rightarrow d\tilde t +dr \frac{v}{1-v^2} \,.
\label{CoordinateTransformation}
\end{equation}
Then the corresponding action contains two terms:
\begin{equation}
S=-E dt \pm \int dr \frac{1}{1-v^2}\sqrt{E^2-m^2(1-v^2)} \,,
\label{SScgh}
\end{equation}

The first term in Eq. (\ref{SScgh})  gives the missing tunneling exponent:
\begin{eqnarray}
 \exp{ \left(-2{\rm Im} \int E dt \right)} = 
\nonumber
  \\
 = \exp{ \left(-2{\rm Im} \int \left(E d\tilde t + dr \frac{Ev}{1-v^2} \right) \right)} =
\nonumber
 \\
 = \exp{ \left(-2{\rm Im} \int dr \frac{Ev}{1-v^2}  \right)} =
\nonumber
 \\
 =  \exp{ \left(-2\pi E R\right)} %=  \exp{ \left(-4\pi EMG\right)} 
 =  \exp{ \left(-\frac{E}{2T_H} \right)} \,.
\label{Et}
\end{eqnarray}

So, the total tunneling exponent is 
\begin{equation}
 \exp{ \left(-2\,{\rm Im}\, S\right)}= \exp{ \left(-\frac{E}{2T_H}\right)}\exp{ \left(\pm\frac{E}{2T_H}\right)}\,.
 \label{TunnSchw}
\end{equation}
The minus sign corresponds to the radiation from the PG black hole considered in the Schwarschild coordinates,  and it gives the Hawking radiation with $T=T_H$.  

In this approach, the Eq.(\ref{TunnSchw}) also contains two contributions, each corresponding to the effective temperature $2T_H$. But in this case  the absorption is $p_{\rm abs} \propto \exp(-E/2T_H)\exp(E/2T_H)$ and the emission   $p_{\rm abs} \propto \exp(-E/2T_H)\exp(-E/2T_H)$, which naturally reflects the detailed balance principle for the black hole.

\section{$2T_H$ problem for white holes}
\label{2THwhitehole}

According to Refs.\cite{Volovik2020a,Volovik2021f,Volovik2021d,Volovik2022}, the hole object can be in three different states:  the PG black hole, the PG white hole and the intermediate state: the neutral fully static object described by Schwarschild coordinates. These three states correspond to different quantum vacua, which are determined by different global coordinate systems. These states can be obtained from each other by singular coordinate transformations between the global coordinate systems. These transformation are given by the same equation (\ref{CoordinateTransformation}), but this equation is now applied to the transformation of macroscopic objects -- black, white and neutral holes. This is distinct from the transformation of the particle spectrum, which we discussed in relation to the Hawking radiation. 

The singular transformations applied to macroscopic objects have been used for calculations of the macroscopic tunneling transitions between these objects and for calculations of the entropy of each macro-object.\cite{Volovik2020a,Volovik2021f,Volovik2021d,Volovik2022}  In this approach the  neutral  object --- the intermediate state between the black and white holes --  has zero entropy. As distinct from the black hole considered either in the Painleve-Gullstrand coordinates or in the static Schwarschild coordinates, the neutral object does not radiate. 

The white hole, which is obtained by the coordinate transformation from the black hole, has the negative entropy and negative temperature. The latter can be also obtained using the methods discussed in Sec. \ref{2THblackhole}. Since in the white hole the shift velocity is opposite to that in the black hole, the first term in Eq.(\ref{SPG}) has the positive sign in the exponent. As a result
one obtains $p_{\rm em} \propto \exp(E/2T_H)\exp(E/2T_H)$ for emission  and $p_{\rm abs} \propto \exp(E/2T_H)\exp(-E/2T_H)$ for absorption. Only the ratio of the two processes is physical. From this ratio $p_{\rm em}/p_{\rm abs} \propto \exp(E/T_H)$ and from the detailed balance principle one obtains that the temperature of the white hole is $T_{\rm WH}=-T_H$. The temperature of the white hole is with minus sign the temperature of the black hole with the same mass. The same relation between the black and white holes is valid for entropy.\cite{Volovik2022}

\section{$2T_H$ problem in the de Sitter spacetime}

Let us now go to the problem of the double-Hawking temperature in de Sitter spacetime. In the dS spacetime, one has $v^2= r^2H^2$, where $H$ is the Hubble parameter, and the cosmological horizon is at $R=1/H$. The same procedure as in Section \ref{2THblackhole} for the PG black hole gives the Hawking radiation in Eq.(\ref{TH2}) with the Hawking temperature $T_H=H/2\pi$.

However, there are several arguments that in addition there are the processes, which are characterized by the local temperature 
is twice the Hawking temperature, $T_{\rm loc}=H/\pi=2T_H$.\cite{Volovik2009b,Volovik2021g,Volovik2022} 
Such temperature is experienced by matter well inside the horizon. This is supported in particular by calculations of the tunneling rate of ionization of atoms in the de Sitter spacetime.\cite{Volovik2009b}

Let us consider an  atom  at the origin, $r = 0$, in the de Sitter spacetime. The electron bounded to an atom absorbs the energy from the gravitational field of the de Sitter background and escapes from the electric potential barrier.  If the ionization potential is much smaller than the electron mass, $H\ll \epsilon_0 \ll m$, one can use
the nonrelativistic quantum mechanics to estimate the tunneling rate through the barrier. The corresponding radial trajectory $p_r(r)$ is obtained from the classical equation $E(p_r, r) = –\epsilon_0$, which is determined by the Doppler shift $p_rv(r)$ with $v(r)=Hr$:
\begin{eqnarray}
-\epsilon_0 = \frac{p^2_r(r)}{2m} + p_rv(r) \,\,, \,\,v(r)=Hr\,,
\label{ElectronTrajectory1}
\end{eqnarray}
\begin{eqnarray}
p_r(r)= -mv(r) + \sqrt{m^2v^2(r) -2m\epsilon_0}\,.
\label{ElectronTrajectory2}
\end{eqnarray}
The integral over the classically forbidden region $0 < r < r_0=\sqrt{2\epsilon_0/mH^2}$ gives the probability of ionization, which looks as  thermal with the double Hawking temperature:
\begin{eqnarray}
 \exp{ \left(-2\,{\rm Im}\, S\right)}= \exp{ \left(-2mH\int_0^{r_o}\sqrt{r_0^2-r^2}\right)}=
 \nonumber
 \\
 = \exp{ \left(-\frac{\pi\epsilon_0}{H}\right)}= \exp{ \left(-\frac{\epsilon_0}{2T_H}\right)}\,.
\label{ProbabilityIonization}
\end{eqnarray}

 The same local temperature describes the process of  the splitting of the composite particle with mass $m$ into two components with $m_1 +m_2>m$, which is also not allowed in the Minkowski vacuum.\cite{Bros2008,Bros2010,Volovik2009b}  The probablity of this process  (for $m\gg T_{\rm H}$):
\begin{eqnarray}
\Gamma(m \rightarrow m_1+m_2) \sim  \exp{\left(-\frac{ m_1+m_2-m}{ 2T_H}\right)}\,.
\label{mto2}
\end{eqnarray}
In particular, for $m_1 =m_2=m$, the decay rate of massive field in the de Sitter spacetime is obtained
\begin{eqnarray}
  \Gamma \sim  \exp{\left(-\frac{m}{2T_H} \right)} \,,
 \label{MassCreation}
 \end{eqnarray}
which is in agreement with Ref. \cite{Jatkar2012}. 

In both cases the tunneling processes are not related to the cosmological horizon. For example, in case of inonization of atom the electron trajectory is well inside the cosmological horizon:
\begin{eqnarray}
r <r_0= \frac{1}{H}\sqrt{\frac{2\epsilon_0}{m}} \ll  \frac{1}{H}\,.
\label{InsideHorizon}
\end{eqnarray}
Another important feature of these processes is that they both violate the de Sitter symmetry, since the atom, which is ionized, or the particle, which is splitted, are external objects. They do not belong to the de Sitter vacuum.

The same phenomenon of local temperature $T_{\rm loc}=2T_H$ can be obtained from consideration of the action (\ref{SPG}). Till now we considered this action for the calculations  of the imaginary part for particle with energy $E>m$ on the trajectory  in the complex plane  which connects the trajectory inside the horizon and the trajectory outside the horizon. This corresponds to the creation of two particles: one inside the horizon and another one outside the horizon. 

However, there is also the trajectory which allows for the creation of a single particle fully inside the cosmological horizon.  In this creation from "nothing", the particle with mass $m$ must have zero energy, $E=0$. This is possible, as follows from the second term in Eq.(\ref{SPG}), which  gives the following imaginary part of the action at $E=0$:
\begin{eqnarray}
 {\rm Im}\, S(E=0)= m \int _0^{1/H}\frac{dr}{\sqrt{1- r^2H^2}} =\frac{\pi}{2}\frac{m}{H}\,.
 \label{LocalCreation1}
 \end{eqnarray}
 The probability of radiation
 \begin{eqnarray}
  \exp{ \left(-2\,{\rm Im}\, S\right)}= \exp{ \left(-\frac{\pi m}{H}\right)}=\exp{ \left(-\frac{m}{2T_H}\right)}\,.
   \label{LocalCreation2}
\end{eqnarray}
again corresponds to the thermal creation of particles by the environment with local  temperature equals to the double Hawking temperature, $T_{\rm loc}=H/\pi=2T_H$. And again, the tunneling trajectory is fully inside the horizon, and this process is possible because the de Sitter symmetry is violated.

\section{Two processes of the black hole radiation}

Two processes -- related and not related to the event horizon --  can be also found in the black hole physics. Let us consider the process of creation of particle with zero energy, $E=0$, without creation of its partner inside the black hole. Observation of such single particle creation is possible, if the detector is at finite distance $R_{\rm o}$ from the black hole. Then the second term in (\ref{SPG}) gives
\begin{eqnarray}
 {\rm Im}\, S(E=0)= m \int _R^{R_{\rm o}}\frac{dr}{\sqrt{1- R/r}} \,.
 \label{SingleBH1}
 \end{eqnarray}
Far from the horizon, at $R_{\rm o}\gg R$, this process is exponentially suppressed: 
\begin{equation}
 p \sim \exp{(-2mR_{\rm o}})\,\,,\,\, R_{\rm o}\gg R \,.
 \label{SingleBH2}
\end{equation}
This process is possible, because the existence of the detector violates the symmetry of the black hole spacetime.

The more general case with nonzero $E<m$ was considered in Ref.\cite{Jannes2011}. It is the  combined process, at which the Hawking radiation is measured by the observer at finite distance $R_{\rm o}$ from the black hole. The process is described by two tunneling exponents. For $R_{\rm o}\gg R$ one obtains:\cite{Jannes2011} 
\begin{eqnarray}
 p \sim  \exp\left(-\frac{E}{T_H}\right),\,\, E>m\,,
 \label{SingleBH3}
 \\
  p \sim  \exp\left(-\frac{E}{T_H}\right) \exp\left(-2R_{\rm o}\sqrt{m^2-E^2}\right), E<m.
 \label{SingleBH4}
\end{eqnarray}
 The first exponent in Eq.(\ref{SingleBH4})  comes from the horizon and corresponds to the conventional Hawking radiation, and the second one describes the process of tunneling of the created particle, which occurs outside the horizon.
As distinct from the de Sitter case, the second process, which takes place only at $E<m$, is exponentially suppressed for $R_{\rm o}\gg R$ and thus does not look as thermal. 

 \section{Acceleration:  Unruh effect vs local processes}

Let us now consider the Unruh effect,\cite{Unruh1976} where one may also expect two independent processes. One of them corresponds to the thermodynamics of  the vacuum in the accelerated frame. The second one corresponds to the radiation experienced by accelerated external object such as atom. Let us consider these two processes in the  Rindler spacetime. For 1+1 case one has 
\begin{eqnarray}
 ds^2=g_{\mu\nu}d^\mu x d^\nu x = (1+ax)^2 dt^2 - dx^2 \,,
 \label{Rindler1}
\\
 g^{\mu\nu} p_\mu p_\nu = \frac{E^2}{(1+ax)^2} -p^2 =m^2\,.
 \label{Rindler1}
\end{eqnarray}

The first process describes the pair creation from the horizon in the Rindler spacetime at $x= -1/a$. It is characterized by Unruh temperature $T_U=a/2\pi$:
\begin{equation}
\exp{ \left(-2 {\rm Im}\int dx \,p(x)\right)}=  \exp{ \left(-\frac{2\pi E}{a} \right)}=  \exp{ \left(-\frac{E}{T_U} \right)}.
 \label{UnruhRadiation}
\end{equation}
One may expect that the ionization of atom is also determined by the thermal bath with Unruh temperature.

However, there is also the local process of ionization of atom, which takes place well inside  the horizon, at $x \ll 1/a$.
The trajectory of the nonrelativistic electron with $\epsilon_0 \ll m$
\begin{eqnarray}
-\epsilon_0 = \frac{p^2(x)}{2m} + ma x\,,
\label{ElectronTrajectoryUnruh}
\end{eqnarray}
 gives the following rate of ionization:\cite{Landau_Lifshitz_3}
\begin{equation}
\exp{ \left(-\frac{4\sqrt{2}}{3}  \frac{\epsilon_0}{a}\left( \frac{\epsilon_0}{m}\right)^{1/2}\right)} \,.
 \label{IonizationUnruh}
\end{equation}
Since $\epsilon_0 \ll m$, the local process of ionization essentially exceeds the rate of thermal ionization $\exp{ \left(-\frac{2\pi \epsilon_0}{a}\right)}$. 

This happens because the existence of the atom violates the symmetry of the Rindler spacetime.
This is very similar to the local process of ionization in the de Sitter spacetime,  where atom violates  the symmetry of the de Sitter spacetime. But now violation of the symmetry is more crucial. As distinct from the de Sitter case, the ionization  does not look as thermal, and thus there is no doubling of Unruh temperature.

 \section{Relation between global and local processes in de Sitter spacetime}

We considered the local processes in the de Sitter, Rindler and Painleve-Gullstrand spacetimes (the combined processes are considered in Ref.\cite{Volovik2022a}). In all three cases, in addition to the processes related to horizons, there are the local processes, which violate the symmetries of these spacetime. The radiation in these local processes differs from by the processes determined by the corresponding Gibbons-Hawking, Unruh and Hawking temperatures. Examples of the local processes are the single-particle creation, and the ionization of the atom. Both do not exist in the pure vacuum states. These processes are possible only due to interaction of the quantum fields in the vacuum with the external object (the detector or an atom). It is the presence of the detector, which  violates the symmetry of the empty spacetime.  Without the detector, the single particle creation is impossible, and only the creation of the pair of entangled particles is allowed, which corresponds to the Hawking and Unruh radiation. 

Among these 3 spacetimes, the de Sitter spacetime is specific, since the local process looks like thermal and the corresponding local temperature is related to the Hawking temperature, $T_{\rm loc}=H/\pi=2T_H$.  This suggests that the Hawking radiation can be represented as the correlated (cotunneling) creation of the entangled pair in the bath with the local temperature $T=2T_H$:
\begin{equation}
\exp{ \left(-\frac{E}{2T_H}\right)}\exp{ \left(-\frac{E}{2T_H}\right)}=\exp{ \left(-\frac{E}{T_H}\right)}\,.
 \label{DoubleHawking}
\end{equation}

This connection between the two processes also suggests that in the presence of the detector (or atom) the symmetry of the de Sitter spacetime partially survives. In the de Sitter Universe all the points in space are equivalent, and the position of the observer may serve as the event horizon for some distant observers.\cite{Volovik2022}
The observer at a given point can see the Hawking radiation as creation of two correlated particles according to Eq.(\ref{DoubleHawking}). On the other hand, the far-distant observer, will see only single particle, which comes from his horizon.
For him it will be the Hawking radiation with Hawking temperature $T_H$.

This symmetry is violated also in the inflationary stage of the expansion of the Universe, which leads to strong deviation from the thermal law except for the limit case $M\ll H$, where the Hawking temperature $T_H$ is obtained.\cite{Starobinsky1986,Starobinsky1994}

 \section{Conclusion}

There is the connection between Eqs. (\ref{LocalCreation2}) and (\ref{DoubleHawking}) for the de Sitter spacetime on one hand and the similar equations (\ref{TH2}) and (\ref{TunnSchw}) for the black hole horizon on the other hand. In both cases the temperature $T=2T_H$ enters. However, the physics is different. 

In case of the PG black hole, there are two contributions to the tunneling process of radiation, each governed by the   temperature $T=2T_H$. They are coherently combined to produce the radiation with the Hawking temperature $T_H$.  This process can be traditionally interpreted as the pair creation of two entangled particles, of which one goes towards the centre of the black hole, while the other one escapes from the black hole.

In case of de Sitter spacetime, the  temperature $T=2T_H$ is physical. Instead of the creation of the entangled pair, this local temperature describes the thermal creation of a single (non-entangled) particle inside the cosmological horizon. The local processes also take place outside of the black hole horizon and inside the Rindler horizon. The local process is highly suppressed in case of the black hole, see Eqs.(\ref{SingleBH2}) and (\ref{SingleBH4}), but is dominating in case of the Rindler spacetime, see Eq.(\ref{IonizationUnruh}).

How the local processes influence the thermodynamics of the de Sitter spacetime is an open question,\cite{Volovik2022}
as well as the problem of the radiation during the de Sitter stage of expansion.\cite{Starobinsky1986,Starobinsky1994,Polyakov2008,Polyakov2012,Akhmedov2014}

  {\bf Acknowledgements}. I thank E. Akhmedov for discussions. This work has been supported by the European Research Council (ERC) under the European Union's Horizon 2020 research and innovation programme (Grant Agreement No. 694248).


\begin{thebibliography}{999}

 \bibitem{Hawking1975}
 S.W. Hawking, 
 Particle creation by black holes,
 Commun. Math. Phys. {\bf 43}, 199-220 (1975).
 
 \bibitem{Volovik1999}
 G.E. Volovik,  
Simulation of Painleve-Gullstrand black hole in thin $^3$He-A film,  
Pis'ma ZhETF {\bf 69}, 662 -- 668 (1999), 
JETP Lett.  {\bf 69}, 705 -- 713 (1999); 
gr-qc/9901077.

\bibitem{Wilczek2000}
M.K. Parikh and F. Wilczek, 
Hawking radiation as tunneling,
Phys. Rev. Lett. {\bf 85}, 5042 (2000).

\bibitem{Volovik2009}
G.E. Volovik,  
Particle decay in de Sitter spacetime via quantum tunneling,
Pis'ma ZhETF {\bf 90}, 3--6 (2009); 
JETP Lett. {\bf 90}, 1--4 (2009);
arXiv:0905.4639 [gr-qc].

\bibitem{Painleve} 
P. Painlev\'e, 
La m\'ecanique classique et la th\'eorie de la relativit\'e, 
 C. R. Acad. Sci. (Paris) {\bf 173} , 677 (1921).
 
 \bibitem{Gullstrand} 
A. Gullstrand,
 Allgemeine L\"osung des statischen Eink\"orper-problems in der Einsteinschen Gravitations-theorie,
Arkiv. Mat. Astron. Fys. {\bf 16}, 1-15 (1922).

\bibitem{Volovik2022b} 
G.E. Volovik,
Painlev\'e-Gullstrand coordinates for Schwarzschild-de Sitter spacetime,
arXiv:2209.02698 [gr-qc].


 \bibitem{Hooft1984} 
G. ’t Hooft, 
On the quantum structure of a black hole,
J. Geom. Phys. {\bf 1}, 45 (1984).

\bibitem{Akhmedov2006}
E.T. Akhmedov, V. Akhmedova, D. Singleton,
Hawking temperature in the tunneling picture,
Phys. Lett. B {\bf 642}, 124--128 (2006). 

\bibitem{Mitra2007}
P. Mitra, 
Phys.Lett. B {\bf 648}, 240--242 (2007),
arXiv:hep-th/0611265v3. 

\bibitem{Laxmi2021}
Y. Onika Laxmi, T. Ibungochouba Singh, I. Ablu Meitei,
Modified entropy of Kerr-de Sitter black hole in Lorentz symmetry violation theory,
arXiv:2112.14545.

\bibitem{Singha2022}
C. Singha, P. Nanda and P. Tripathy,
Hawking radiation in multi-horizon spacetimes,
arXiv:2206.06433.

\bibitem{Srinivasan1999}
K. Srinivasan and T. Padmanabhan,
Particle production and complex path analysis,
Phys. Rev. D {\bf 60}, 024007 (1999).

 \bibitem{Volovik2020a}
 G.E. Volovik,
Varying Newton constant and black hole to white hole quantum tunneling,
MDPI, Universe {\bf 6}, 133 (2020),
arXiv:2003.10331.

\bibitem{Volovik2021f}
G.E. Volovik,
From black hole to white hole  via the intermediate static state,
Modern Physics Letters A {\bf 36}, 2150117  (2021),
arXiv:2103.10954.

\bibitem{Volovik2021d}
G.E. Volovik,
Effect of the inner horizon on the black hole thermodynamics: Reissner-Nordstr\"om black hole and Kerr black hole,
Mod. Phys. Lett. A  {\bf 36}, 2150177 (2021).

\bibitem{Volovik2022}
G.E. Volovik,
Macroscopic quantum tunneling: from quantum vortices to black holes and Universe,
ZhETF {\bf 162}, 449--454 (2022), 
JETP {\bf 135}, 388--408 (2022),
arXiv:2108.00419.

\bibitem{Volovik2009b}
G.E. Volovik, 
Particle decay in de Sitter spacetime via quantum tunneling, 
JETP Lett. {\bf 90},  1--4 (2009); 
arXiv:0905.4639 [gr-qc].

\bibitem{Volovik2021g} 
G.E. Volovik,
Double Hawking temperature in de Sitter Universe and cosmological constant problem,
arXiv:2007.05988.

\bibitem{Bros2008} 
J. Bros, H. Epstein, and U. Moschella, 
Lifetime of a massive particle in a de Sitter universe,
JCAP 0802:003 (2008); 
arXiv:hep-th/0612184.

\bibitem{Bros2010} 
J. Bros, H. Epstein, M. Gaudin, U. Moschella and V. Pasquier,
Triangular invariants, three-point functions and particle stability on the de Sitter universe,
Commun. Math. Phys. {\bf 295}, 261--288 (2010).

\bibitem{Jatkar2012} 
D.P. Jatkar, L. Leblond and A. Rajaraman,
Decay of massive fields in de Sitter space,
Phys. Rev. D {\bf 85}, 024047 (2012). 

\bibitem{Jannes2011}
G. Jannes,
Hawking radiation of $E<m$ massive particles in the tunneling formalism,
JETP Lett. {\bf 94},  18--21  (2011).

\bibitem{Unruh1976}
W.G. Unruh, 
Notes on black-hole evaporation,
Phys. Rev. D {\bf 14}, 870 (1976).

\bibitem{Landau_Lifshitz_3}
 L.D. Landau  and  E.M. Lifshitz, 
 Course of Theoretical Physics, Volume 3, 
Quantum Mechanics.

\bibitem{Volovik2022a} 
G.E. Volovik,
Particle creation: Schwinger + Unruh + Hawking,
Pis’ma v ZhETF {\bf 116}, 577 (2022),
JETP Lett. {\bf 116},   (2022),
DOI: 10.1134/S0021364022601968,
arXiv:2206.02799.

\bibitem{Starobinsky1986}
A. Starobinsky, 
Stochastic de Sitter (inflationary) stages in the Early Universe, 
Lect. Notes in Physics {\bf 246}, 107--126 (1986).

\bibitem{Starobinsky1994}
 A. A. Starobinsky and  J. Yokoyama, 
 Equilibrium state of a self-interacting scalar field in the de Sitter background, 
 Phys. Rev. D {\bf 50}, 6357 (1994) [arXiv:9407016],

\bibitem{Polyakov2008}
A.M. Polyakov, 
De Sitter space and eternity, 
Nucl. Phys. B {\bf 797}, 199--217 (2008);
arXiv:0709.2899

\bibitem{Polyakov2012}
A.M. Polyakov, 
Infrared instability of the de Sitter space,
arXiv:1209.4135 [hep-th].

\bibitem{Akhmedov2014}
E.T. Akhmedov,
Lecture notes on interacting quantum fields in de Sitter space,
International Journal of Modern Physics {\bf 23}, 1430001 (2014).

\end{thebibliography}
\end{document}